\documentclass[superscriptaddress,preprint,aps,showpacs]{revtex4}%
\usepackage{amsfonts}
\usepackage{amsmath}
\usepackage{amssymb}
\usepackage{graphicx}
\usepackage{xcolor}%
\begin{document}
\title{Fractional quantum Hall effect in topological insulators: The role of Zeeman effect}
\author{Zhigang Wang}
\affiliation{LCP, Institute of Applied Physics and Computational Mathematics, P. O. Box
8009, Beijing 100088, China}
\author{Fawei Zheng}
\affiliation{LCP, Institute of Applied Physics and Computational Mathematics, P. O. Box
8009, Beijing 100088, China}
\author{Zhen-Guo Fu}
\affiliation{Beijing Computational Science Research Center, Beijing 100089, China}
\author{Ping Zhang}
\thanks{Corresponding author. Email address: zhang\_ping@iapcm.ac.cn}
\affiliation{LCP, Institute of Applied Physics and Computational Mathematics, P. O. Box
8009, Beijing 100088, China}
\affiliation{Beijing Computational Science Research Center, Beijing 100089, China}

\begin{abstract}
We study the role of Zeeman effect in fractional quantum Hall effect (FQHE) on
the surface of topological insulators (TIs). We show that the effective
pseudopotentials of the Coulomb interaction are reformed due to Zeeman effect,
which are quite different from those in graphene. By exactly diagonalizing the
many-body Hamiltonian in the sphere geometry, we find that the ground state
energies and the excitation gaps at $\nu$=$1/3$ FQHE between the $n$=$\pm1$
Landau levels (LLs) render asymmetry, and the FQHE state at the $n$=$1$ LL is
more robust than that at $n$=$-1$ LL since the excitation gap at $n$=$1$ LL is
larger than that at $n$=$-1$ LL.

\end{abstract}

\pacs{73.43.-f, 73.25.+i, 71.10.-w}
\maketitle

\section{ Introduction}

Because of the rich physical properties in fractional quantum Hall effect
(FQHE), it has proved to be a vital and exciting subfield of condensed-matter
physics to the present day. The electronic state giving rise to the
experimentally \cite{Tsui,Chang,Kawaji, Chang1,Ebert} observed FQHE features
should be of a many-particle origin, and has been studied intensively with
various theoretical\ tools \cite{Laughlin, Halperin, Haldane, Yoshioka,
MacDonald, Morf, Chakraborty, Girvin1990, Das1997, Murthy,Jain2007}. At
present, most of experiments on this issue are performed in high-mobility
GaAs-based semiconductor structures. However, it is a still a challenge to
detect the fractional quantum Hall state with high filling factors \cite{Dean}
(such as $\nu$=$5/2$) due to the restrict of carrier mobility and the weak
electron interactions in two dimensional semiconductor materials.

Fortunately, recently discovered two-dimensional electron systems, including
graphene \cite{CastroNeto, Goerbig1}, two-dimensional HgTe quantum wells
\cite{Bernevig,Konig}, and surface state of three-dimensional topological
insulators (TI) \cite{Hasan,QiXL}, which possess linear dispersion relation
near the Dirac points with high Fermi velocity, may be good platforms to
detect the FQHE. The unusual integer quantum Hall effect (IQHE)
\cite{Novoselov, Zhang} with the $4e^{2}/h$ separation in the quantized Hall
conductivity has been found experimentally in graphene, where the Landau
levels (LLs) are four-fold degenerate without Zeeman effect and weak
spin-orbit couplings. Soon after this discovery, Nomura and MacDonald
\cite{MacDonald2006} suggested the interaction-driven quantum Hall effects
with intermediate integer value of $e^{2}/h$ because of the charge gaps due to
the electron-electron interaction. FQHE in graphene was also predicted
theoretically \cite{Toke, Apalkov, Goerbig}, and the $1/3$\texttt{-}FQHE state
was later measured \cite{Abanin,Ghahari} successfully in suspended graphene.

Because of the Dirac nature of the electrons, many properties of the
two-dimensional and three-dimensional TIs are similar to those of graphene.
Recently, the graphene-like IQHE has been observed in strained bulk HgTe
\cite{Brune}\ and the LL spectrum has been measured in Bi$_{2}$Se$_{3}$ by
using scanning tunneling microscopy \cite{Cheng, Hanaguri}. Features in the
Hall resistance at fractional filling factors have been speculated to be
related to the FQHE of TIs \cite{Analytis, Xiong}.

However, the finite thickness of TI sample in the growth direction will affect
the nature of the FQHE states on TI surfaces \cite{Apalkov2}, which is in
contrast to the case in graphene. In addition, there exists an another
remarkable difference between graphene and TI, among which topologically
protected Dirac cones on the surface of three-dimensional TI arise from the
real spin-orbit coupling while the Dirac cone in graphene results from the
pseudospin-orbit coupling. As a consequence, the external magnetic field
induced Zeeman effect in TI and graphene are quite different. The Zeeman
effect reforms the LLs as $sgn(n)\sqrt{\Delta_{Z}^{2}\mathtt{+}|n|\left(
\hbar\omega\right)  ^{2}}$ in TI ($\Delta_{Z}$ is the Zeeman energy, $n$ is
the LL index and $\omega$=$eB/mc$ is the cyclotron frequency), while in
graphene the LLs split as $\mathtt{\pm}\sqrt{|n|}\hbar\omega\mathtt{\pm}%
\Delta_{Z}$ in the presence of Zeeman effect. Thereby, the influence of Zeeman
effect on FQHE in TI should be quite different from that in graphene.
Furthermore, on one side, the Zeeman term can perturb the Landau quantization
in TI materials for a sufficiently large g-factor, which can reach
$\mathtt{\sim}$30 in recent experiment performed on the surface of Bi$_{2}%
$Se$_{3}$ \cite{Analytis}, in contrast, the g-factor in graphene is very small. On the other side, sufficiently strong magnetic field
as large as hundreds of tesla, which can also bring on prominent Zeeman
effect, is achievable in current experimental capabilities.

Therefore, because of its importance both from basic point of interest and to
the analysis of unconventional properties of TI-based FQHE, in the present
paper we address the role of Zeeman effect in FQHE of the massless Dirac
electrons on TI surface by presenting an attempt at the theoretical evaluation
of the effective pseudopotentials of the electron-electron interactions and
the ground (excitation) state nature at $1/3$-FQHE between the $n$=$\pm1$ LLs
problems. We find the following facts: (i) The effective pseudopotentials of
the Coulomb interaction $V_{\text{eff}}^{(n,m)}$ ($m$ is the relative angular
momentum of two electrons) are reformed due to the Zeeman term. At the first
few relative angular momentum $m$, we have $V_{\text{eff}}^{(|n|,m)}%
\mathtt{>}V_{\text{eff}}^{(-|n|,m)}$ for any LL $n\mathtt{\neq}0$, while by
increasing $m$ to a moderate value, an inverse of the effective
pseudopotentials happens $V_{\text{eff}}^{(|n|,m)}\mathtt{<}V_{\text{eff}%
}^{(-|n|,m)}$. (ii) By taking into account the Zeeman effect and exactly
diagonalizing the many-body Hamiltonian in the sphere geometry, the ground
state energies and the excitation gaps at $\nu$=$1/3$ FQHE between the
$n$=$\pm1$ LLs exhibit asymmetry. (iii) Comparing with that at the $n$=$-1$
LL, the FQHE state at the $n$=$1$ LL is more robust because the excitation gap
at $n$=$1$ LL is larger than that in $n$=$-1$ LL in the presence of the Zeeman
energy. These findings obtained can not be observed in graphene within Zeeman
effect, because the Zeeman energies in graphene just translate the LLs by $\pm
g\mu_{B}B$, but induce no influence on the the Haldane's pseudopotential of
electron-electron interactions.

\section{ Model and Method}

We start from the free electron low-energy effective Hamiltonian
of TI surface in the presence of a perpendicular magnetic field, which is
given by%
\begin{equation}
H=v_{F}\boldsymbol{\sigma}\cdot\mathbf{\Pi}+g\mu_{B}B\sigma_{z},\label{1}%
\end{equation}
where $v_{F}$ is the Fermi velocity and $\boldsymbol{\Pi}$=$\boldsymbol{p}%
\mathtt{+}e\boldsymbol{A}/c$ is the two-dimensional canonical momentum. It is
easy to obtain the eigenstates of Eq. (\ref{1}), by choosing symmetric gauge
$\boldsymbol{A}$=$B\left(  -y/2,x/2,0\right)  $ and introducing the ladder
operators $a^{\dag}\mathtt{=}\frac{1}{\sqrt{2}}\left(  \frac{\bar{z}}%
{2l}\mathtt{-}2l\partial_{z}\right)  $,\ $a\mathtt{=}\frac{1}{\sqrt{2}}\left(
\frac{\bar{z}}{2l}\mathtt{+}2l\partial_{z}\right)  $, as
\begin{equation}
\Psi_{n,m}(r)=\left(
\begin{array}
[c]{c}%
\alpha_{n}\phi_{|n|-1,m}\\
\beta_{n}\phi_{|n|,m}%
\end{array}
\right)  ,\label{2}%
\end{equation}
where $\phi_{nm}$ is the eigenstate of the 2D Hamiltonian with
non-relativistic quadratic dispersion relation in $n$th LL with angular
momentum $m$. Here, $z$=$x$+$iy$, and the magnetic length $l$=$\sqrt{eB/\hbar
c}$, which is the length unit throughout this paper. $\alpha_{n}%
\mathtt{=}\mathtt{-}i$sgn$(n)\frac{\cos\varphi_{n}}{\sqrt{2\left(
1\mathtt{-}\text{sgn}(n)\sin\varphi_{n}\right)  }}$ and $\beta_{n}%
\mathtt{=}\sqrt{\frac{\left(  1\mathtt{-}\text{sgn}(n)\sin\varphi_{n}\right)
}{2}}$ for $n\mathtt{\neq}0$, while $\alpha_{n}$=$0$ and $\beta_{n}$=$1$ for
$n\mathtt{=}0$ LL, which contain the information of Zeeman effect since
$\varphi_{n}$=$g\mu_{B}B/(\hbar v_{F}l^{-1}\sqrt{2|n|})\mathtt{\equiv}%
\varphi/\sqrt{2|n|}$. The corresponding eigenenergies are given by $E_{n\neq
0}\mathtt{=}$sgn$(n)\sqrt{\left(  g\mu_{B}B\right)  ^{2}\mathtt{+}%
2|n|\hbar^{2}v_{F}^{2}l^{-2}}$, and $E_{n=0}$=$-g\mu_{B}B$, which are
different from LLs in graphene $E_{n}^{\text{gra}}\mathtt{=}\mathtt{\pm}%
g\mu_{B}B\mathtt{\pm}\sqrt{2|n|}\hbar v_{F}l^{-1}$ taken into account the
Zeeman splitting in the effective Hamiltonian at the Dirac point $K$ (or
$K^{\prime}$) of graphene $H_{0}^{\text{gra}}\mathtt{=}v_{F}\boldsymbol{\sigma
}\mathtt{\cdot}\boldsymbol{\Pi}\mathtt{\otimes}I_{2\times2}\mathtt{+}%
I_{2\times2}\mathtt{\otimes}g\mu_{B}\boldsymbol{B}\mathtt{\cdot}%
\boldsymbol{s}$ with $\boldsymbol{\sigma}$ the pseudospin and $\boldsymbol{s}$
the real spin.

In order to investigate the properties of TI surface FQHE at partially
occupied LLs with fractional filling factors, we should firstly consider
Coulomb interaction $V\left(  \boldsymbol{r}\right)  \mathtt{=}e^{2}/\epsilon
r$ between two electrons in $n$th LL with relative angular momentum $m$. It is
expressed as the Haldane's pseudopotential
\begin{align}
V_{\mathtt{eff}}^{(n,m)}  &  \mathtt{=}\left\langle \left\langle nm_{1}%
;nm_{2}\right\vert \right\vert V\left\vert \left\vert nm_{3};nm_{4}%
\right\rangle \right\rangle \nonumber\\
&  \mathtt{=}\frac{\cos^{4}\varphi_{n}}{4\left(  1\mathtt{-}\text{sgn}%
(n)\sin\varphi_{n}\right)  ^{2}}V_{m}^{(|n|-1)}\text{+}\frac{\cos^{2}%
\varphi_{n}}{2}V_{m}^{(|n|,|n|-1)}\nonumber\\
&  \mathtt{+}\frac{\left(  1\mathtt{-}\text{sgn}(n)\sin\varphi_{n}\right)
^{2}}{4}V_{m}^{(|n|)}, \label{3}%
\end{align}
where $\left\vert \left\vert nm_{1};nm_{1}\right\rangle \right\rangle
\mathtt{=}\Psi_{n,m_{1}}\mathtt{\otimes}\Psi_{n,m_{2}}$ and
\begin{align}
V_{m}^{(|n|)}  &  =\int\frac{d^{2}k}{\left(  2\pi\right)  ^{2}}\frac{2\pi}%
{k}\left[  L_{|n|}\left(  k^{2}/2\right)  \right]  ^{2}L_{m}\left(
k^{2}\right)  e^{-k^{2}},\label{4}\\
V_{m}^{(|n|,|n|-1)}  &  =\int\frac{d^{2}k}{\left(  2\pi\right)  ^{2}}%
\frac{2\pi}{k}L_{|n|}\left(  k^{2}/2\right) \nonumber\\
&  \times L_{|n|-1}\left(  k^{2}/2\right)  L_{m}\left(  k^{2}\right)
e^{-k^{2}}, \label{5}%
\end{align}
with $L_{m}\left(  x\right)  $ the Lagueree polynomials. In what follows, the
energy unit is chosen as $\varepsilon_{c}$=$e^{2}/\epsilon l$, and focus on
the role of Zeeman effect in TI surface FQHE. It is clear that the Zeeman term
can perturb the Landau quantization itself for a sufficiently large g-factor
or a sufficiently strong external magnetic field $B$.\begin{figure}[ptb]
\begin{center}
\includegraphics[width=0.5\linewidth]{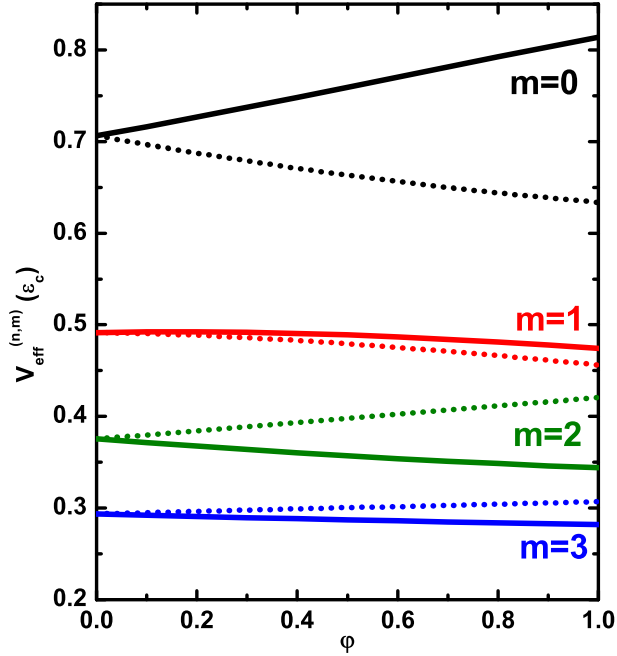}
\end{center}
\caption{(Color online) The effective pseudopotentials of the Coulomb
interaction $V_{\mathtt{eff}}^{(n,m)}$ between two electrons at $n$=$\pm1$ LLs
with the lowest four relative angular momentum $m$ as a function of Zeeman
parameter $\varphi$=$g\mu_{B}B/\hbar v_{F}l^{-1}$. The solid and dotted lines
correspond to $n$=$1$ and $n$=$-1$ LLs, respectively. }%
\label{fig1}%
\end{figure}

\section{ Results and Discussions}

One can easily find from Eq. (\ref{3}) that the effective
pseudopotential at the $n$=$0$ LL is independent on the Zeeman energy. The
effective pseudopotentials at other LLs ($n\mathtt{\neq}0$), however, are
dependent on the Zeeman energy. Because the stability of FQHE phenomenon can
only occur at $n$=$0,\pm1$ LLs on a TI surface \cite{DaSilva}, we only plot in
Fig. \ref{fig1} the effective pseudopotentials of the Coulomb interaction
$V_{\mathtt{eff}}^{(n,m)}$ between two electrons at $n$=$\pm1$ LLs with the
lowest four relative angular momentum $m$=$0,\mathtt{\cdots},3$ as a function
of Zeeman parameter $\varphi$. We notice that on one side, the
pseudopotentials $V_{\mathtt{eff}}^{(1,m)}$ and $V_{\mathtt{eff}}^{(-1,m)}$
are degenerate in the absence of Zeeman term ($\varphi$=$0$), which could also
be straightforwardly observed from Eq. (\ref{3}) by setting $\varphi_{n}$=$0$.
On the other side, the degeneracy between $V_{\mathtt{eff}}^{(1,m)}$ and
$V_{\mathtt{eff}}^{(-1,m)}$ is removed with increasing the Zeeman parameter
$\varphi$. Interestingly, one can see in Fig. \ref{fig1} that under the same
Zeeman parameter $\varphi$ (for example, $\varphi$=$0.2$) the splitting of
pseudopotentials for relative angular momentum $m$=$1$, i.e., $\left\vert
V_{\mathtt{eff}}^{(1,1)}\mathtt{-}V_{\mathtt{eff}}^{(-1,1)}\right\vert $ is
smaller than that for other lower relative angular momentum $m$=$0,2,$and $3$.
We also find in Fig. \ref{fig1} that, for the electrons at LL $n$=$\pm1$ the
pseudopotentials $V_{\mathtt{eff}}^{(1,m)}$%
$>$%
$V_{\mathtt{eff}}^{(-1,m)}$ only for relative angular momentum $m$=$0$ and
$1$, but $V_{\mathtt{eff}}^{(1,m)}\mathtt{<}V_{\mathtt{eff}}^{(-1,m)}$ for
higher angular momentum $m$.

\begin{figure}[ptb]
\begin{center}
\includegraphics[width=0.6\linewidth]{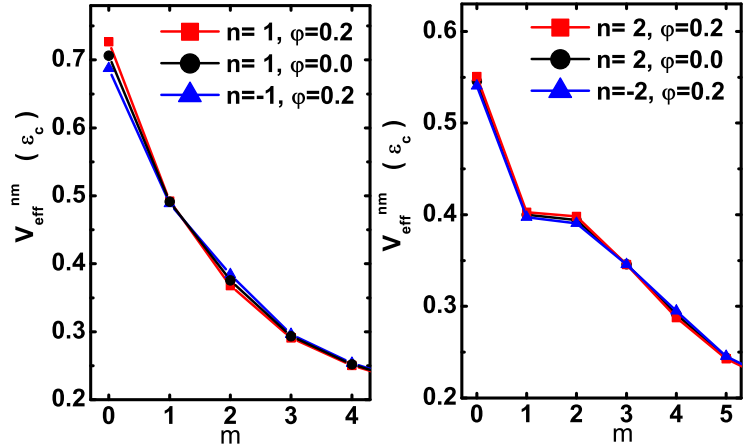}
\end{center}
\caption{(Color online) The effective pseudopotentials $V_{\mathtt{eff}%
}^{(n,m)}$ between two electrons at (a) $n$=$\pm1$ and (b) $n$=$\pm2$ LLs as
functions of the relative angular momentum $m$ with different Zeeman
parameters $\varphi$=$0$ and $\varphi$=$0.2$. The lines are a guide to the
eye.}%
\label{fig2}%
\end{figure}

These results could also be seen from Fig. \ref{fig2}(a), where the red
squares (blue triangles) correspond to $n$=$1$ ($n$=$-1$) with $\varphi$%
=$0.2$. In fact, it is a truth that at the first few relative angular momentum
$m$, we have $V_{\text{eff}}^{(|n|,m)}\mathtt{>}V_{\text{eff}}^{(-|n|,m)}$ for
any LL $n\mathtt{\neq}0$, while by increasing $m$ to a moderate value, an
inverse of the effective pseudopotentials happens $V_{\text{eff}}%
^{(|n|,m)}\mathtt{<}V_{\text{eff}}^{(-|n|,m)}$. As an another example, we also
show the pseudopotentials at LL $n\mathtt{=}\pm2$ verse relative angular
momentum $m$ in Fig. \ref{fig2}(b).

Based on the pseudopotentials for Dirac fermions on the surface of
TIs, one can further obtain the energy spectra of the many-body states at
fractional fillings of the LL by numerically diagonalizing the many-body
Hamiltonian in the spherical geometry \cite{Haldane}. In this point of view,
thus the external magnetic field perpendicular to the surface of sample is
equivalent to a fictitious radial magnetic field produced by a magnetic
monopole at the center of sphere of radius $R$, which is related to the
magnetic fluxes $2S$ through the sphere. It is easy to demonstrate that in the
units of flux quanta $R$=$\sqrt{S}l$, and the many-body states could be
described by the total angular momentum $L$ and its $z$ component $L_{z}$,
while the energy just depends on $L$. In what follows, we investigate the
system with the fractional filling factor $\nu$=$1/\left(  2p\mathtt{+}%
1\right)  $, where $p$ is an integer, especially the $p$=$1$ case, i.e., the
$\nu$=$1/3$ FQHE state which is realized at $S$=$\frac{3}{2}\left(
N\mathtt{-}1\right)  $ in the spherical geometry, where $N$ is the electron number.

\begin{figure}[ptb]
\begin{center}
\includegraphics[width=0.6\linewidth]{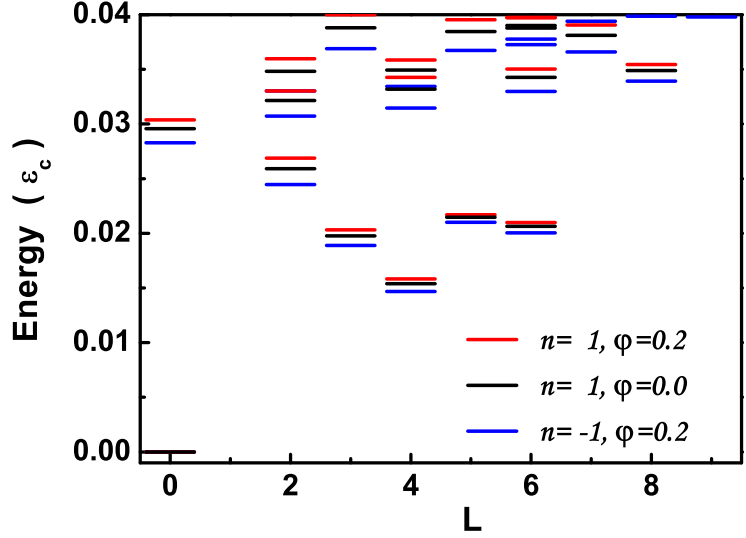}
\end{center}
\caption{(Color online) Exact energy per electron versus the angular momentum
$L$ for $N$=$6$ electrons at $\nu$=$1/3$ FQHE state. The blue and red lines
correspond to the $n$=$1$ and $n$=$-1$ LLs, respectively. The Zeeman parameter
is chosen as $\varphi$=$0.2$. }%
\label{fig3}%
\end{figure}

The typical results of exactly diagonalized energy spectra per
electron for the $\nu$=$1/3$\texttt{-}FQHE state at $n$=$\pm1$ LLs with the
same Zeeman parameter $\varphi$=$0.2$ as well as the degenerate energy spectra
for the $\nu$=$1/3$-FQHE state at $n$=$\pm1$ without Zeeman term are shown in
Fig. \ref{fig3}. One can clearly observe in Fig. \ref{fig3} that the
degenerate energy spectra (black) at $\varphi$=$0$ split into two-class alike
energy spectra when the Zeeman term are introduced. The excited energy spectra
(red) at $n$=$1$ LL are promoted while those (blue) at $n$=$-1$ LL are
degraded. From Fig. \ref{fig3} one can also find two prominent features
brought by the Zeeman term as follows: (i) The FQHE state at the $n$=$1$ LL is
more robust since the excitation gap at $n$=$1$ LL is larger than that at
$n$=$-1$ LL in the presence of the Zeeman term; (ii) However, the promotion
(degradation) of each excited energy is not same with each other, which
indicates that the Zeeman term leads to the asymmetry of the LLs. These two
properties of the Zeeman effect on the FQHE in TIs are unique, which can not
be observed in the popular studied graphene system because the Zeeman energy
in graphene Hamiltonian $H_{0}^{\text{gra}}$ does not affect the Haldane's
pseudopotential but just translate the LLs by $\pm g\mu_{B}B$.
\begin{figure}[ptb]
\begin{center}
\includegraphics[width=0.6\linewidth]{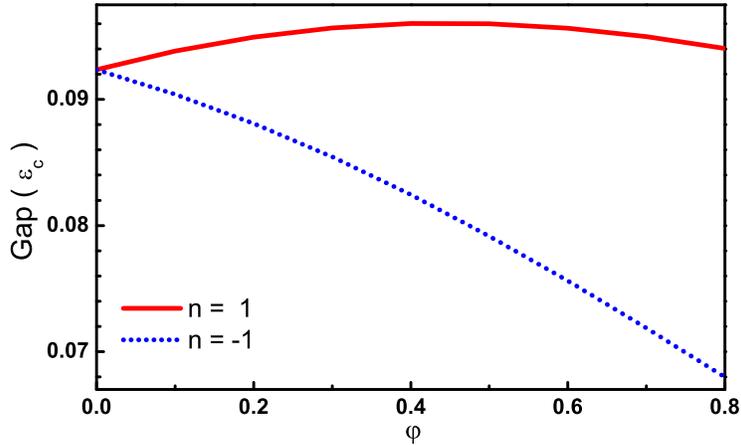}
\end{center}
\caption{(Color online) Excited energy gap versus Zeeman parameter $\varphi$
for $N$=$6$ electrons at $\nu$=$1/3$ FQHE state. The red and blue lines
correspond to the $n$=$1$ and $n$=$-1$ LLs, respectively. }%
\label{fig4}%
\end{figure}

We now turn to study the dependence of the excited gap on the Zeeman parameter
$\varphi$. In Fig. \ref{fig4} we show the excited energy gap (in unit of
$\varepsilon_{c}$) versus Zeeman parameter $\varphi$ for $N$=$6$ electrons at
$1/3$\texttt{-}FQHE state, whose corresponding lowest excited state angular
momentum is $L$=$4$. One can clearly see that the excited gap for $n$=$1$ LL
is always larger than that for $n$=\texttt{-}$1$ LL, which is consistent with
the results in Fig. \ref{fig3}. Besides, the excited gap for $n$=\texttt{-}$1$
LL monotonically decreases with increasing the Zeeman parameter (see the blue
line in Fig. \ref{fig4}). Whereas, the variance of the excited gap for $n$=$1$
LL (the red line in Fig. \ref{fig4}) is weak with increasing the Zeeman
parameter, and it gradually increases to a maximum at $\varphi\mathtt{\approx
}0.42$ and then tends to decrease.

\begin{figure}[ptb]
\begin{center}
\includegraphics[width=0.6\linewidth]{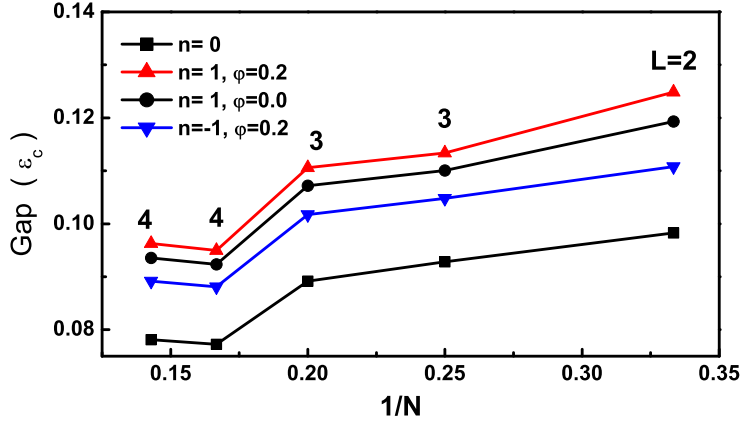}
\end{center}
\caption{(Color online) Excited energy gap versus the inverse of the electron
number $N$ for the $n$=$\pm1$ LLs with different Zeeman parameters $\varphi
$=$0.2$ and $\varphi$=$0$. The integers denote the angular momentums $L$ of
the lowest excited energy states. }%
\label{fig5}%
\end{figure}

The above discussions focus on the system of $N$=$6$ electrons, but the
properties obtained herein are conserved when the electron number $N$ is
changed. The results of the excited gap as a function of $N$ are shown in Fig.
\ref{fig5}. We should point out that, firstly, for the $n$=$0$ LL, the Zeeman
term does not affect the FQHE in TIs and therefore the excited gaps are same
for different values of $\varphi$. Secondly, the case of $n$=$0$ LL is quite
different from that for $n\mathtt{\neq}0$ LLs. From Fig. \ref{fig5} we see
that the excited gap is promoted (degraded) for $n$=$1$ ($n$=$-1$) LL due to
the Zeeman effect. Thirdly, the angular momentum $L$ of lowest excited state
alters with the electron number $N$, which is indicated by the integers in
Fig. \ref{fig5}. For instance, the angular momentum of the lowest excited
states for systems of $N$=$4$ and $5$ electrons are $L$=$3$, while this
becomes $L$=$4$ for systems of $N$=$6$ and $7$ electrons. We also studied the
FQHE with the fractional filling factor $\nu$=$1/\left(  2p+1\right)  $, where
$p$=$2$, $3$,$\cdots$, and obtain the same results as that with $\nu$=$1/3$.

\section{Summary}

In summary, we have studied the Zeeman effect on FQHE in TIs. Compared to
previously studied graphene system, the combination of the real spin-orbit
interaction and the Zeeman term bring on unique physical properties in TIs. We
have found that the effective pseudopotentials of the Coulomb interaction are
reformed owing to the Zeeman effect. By exactly diagonalizing the many-body
Hamiltonian in the sphere geometry, we also found that the ground state
energies and the excitation gaps at $\nu$=$1/3$ FQHE between the $n$=$\pm1$
Landau levels (LLs) exhibit asymmetry. Comparing with the $n$=$-1$ LL, the
FQHE state at the $n$=$1$ LL is more robust since its excitation gap is
larger. We hope our predictions can be confirmed in future experiments.

We thank Tapash Chakraborty and Jize Zhao for helpful discussions about the
pseudopotentials. This work was supported by Natural Science Foundation of
China under Grants No. 90921003, No. 11274049, and No. 11004013, and by the
National Basic Research Program of China (973 Program) under Grant No. 2009CB929103.

\end{document}